\documentclass[a4paper, 10pt, conference]{ieeeconf}

\usepackage[pdftex]{graphicx}
\usepackage{amsmath,amssymb,amsfonts}
\usepackage{url}

\newtheorem{definition}{Definition}
\title{\Large \bf
Observability of Nonlinear Complex Networks in the Presence of Symmetries:\\ A Graphical Approach}

\author{Afroza Shirin$^{1}$, Dionicio F. Rios$^{1}$, and Francesco Sorrentino$^{1}$%
\thanks{*We gratefully acknowledge support from the National Science Foundation through NSF grant CMMI- 1400193, NSF grant CRISP- 1541148 and the  Office of Naval Research  through ONR Award No. N00014-16-1-2637.}
\thanks{$^{1}$Department of Mechanical Engineering, The University of New Mexico, Albuquerque, New Mexico, 87131
{\tt\small ashirin@unm.edu}}%
}

\begin{document}
\maketitle
\thispagestyle{empty}
\pagestyle{empty}

\begin{abstract} Reconstructing the states of the nodes of a dynamical network is a problem of fundamental importance in the study of neuronal and genetic networks. An underlying related problem is that of observability, i.e.,  identifying the conditions under which such a reconstruction is possible. In this paper  we study observability of complex dynamical networks,
where, we consider the effects of network symmetries on observability. We present an efficient algorithm that returns a  minimal set of necessary sensor nodes for observability  in the presence of symmetries.
\end{abstract}

\section{\textbf{Introduction}}

Observability is the ability to deduce the state variables of a system for which only the outputs, often coinciding with a subset of the states, can be measured. There are different methods that can be used to determine observability, ranging from an algebraic test \cite{datta2004numerical} to a graphical approach (GA) \cite{diop1991nonlinear}. The last allows to determine a set of output states from which one can infer information about the rest of the states by analyzing the properties of an inference graph. In \cite{liu2013observability}, this method was applied to complex  networks, for which the ability to directly sense all of the nodes is typically unavailable. Examples of interest are genetic and neural networks, which could be reconstructed by observing a limited number of output states if these provide observability.

In biological networks, such as  genetic or  neuronal networks, an important problem is that of reconstructing the network architecture and the time evolution of the connections from existing data. For these applications, a typical  limitation is the cost and the possibility of sensing the states of the individual nodes. The network inference problem and its application to the genome \cite{de2002linking,friedman2004inferring,de2006unravelling,faith2007large,marbach2009replaying} has received significative attention over the last decade. References \cite{de2010advantages,marbach2010revealing} review and compare the more popular existing approaches. An important question that is relevant to both genetic and neuronal networks, is what is the minimal amount of information needed to be able to reconstruct the state of these networks.

In this paper we will be concerned with the conditions for observability of a complex dynamical network, which provides a \textit{conditio sine qua non} for implementing any type of estimation strategy.

\section{\textbf{Definitions}}

\begin{definition}{\textit{Observability of Non-Linear System.}}
Consider a nonlinear system represented by,
\begin{equation}\label{NLsys}
\dot{\textbf{x}}=\textbf{f}(\textbf{x},\textbf{u}),
\end{equation}where $\textbf{x} \in \mathbb{R}^N$ is the system state, $\textbf{u} \in \mathbb{R}^P$ is the system input, $\textbf{f}:\mathbb{R}^N \times \mathbb{R}^P \rightarrow \mathbb{R}^N$ and
\begin{equation}
\textbf{y}=\textbf{h}(\textbf{x})=
\begin{bmatrix}
h_1(\textbf{x})\\
h_2(\textbf{x})\\
...\\
h_M(\textbf{x})\\
\end{bmatrix}
\end{equation}
is the system output, $\textbf{y} \in \mathbb{R}^M$, $h_i:\mathbb{R}^N \rightarrow \mathbb{R}$. If the initial-state vector $\textbf{x}(0)$ can be found from $\textbf{u}(t)$ and $\textbf{y}(t)$ measured over a finite interval of time, the system in Eq.\ \eqref{NLsys} is said to be \emph{observable}.

The Lie derivative of $\textbf{h}$ with respect to $\textbf{f}$  is defined to be
\begin{equation}
L_\textbf{f}\textbf{h}=
\frac{\partial \textbf{h}}{\partial \textbf{x}}
\textbf{f},
\end{equation}where the function $\textbf{f}$ is of the form:
\begin{equation}
\textbf{f}=
\begin{bmatrix}
f_1(\textbf{x})\\
...\\
f_N(\textbf{x})\\
\end{bmatrix}.
\end{equation}Then the Lie derivative is:
\begin{equation}
L_\textbf{f}\textbf{h}=
\begin{bmatrix}
\frac{\partial \textbf{h}}{\partial x_1},...,\frac{\partial \textbf{h}}{\partial x_N}\\
\end{bmatrix}
\begin{bmatrix}
f_1(\textbf{x})\\
...\\
f_N(\textbf{x})\\
\end{bmatrix}
\end{equation} and the result of this derivative operation is a scalar.

Two states $\textbf{x}_0$ and $\textbf{x}_1$ are distinguishable if there exists an input function $\textbf{u}^*$ such that:
\begin{equation}
\textbf{y}(t,\textbf{x}_0)\neq \textbf{y}(t,\textbf{x}_1),
\end{equation}
where $\textbf{y}(t,\textbf{a}) \equiv \textbf{y}(t)$ for $\textbf{x}(0)=\textbf{a}$. This implies that the system is locally observable about $\textbf{x}_0$  if there exists a neighborhood of $\textbf{x}_0$ such that every $\textbf{x}$ in that neighborhood other than $\textbf{x}_0$ is distinguishable from $\textbf{x}_1$.

A test for local observability about $\textbf{x}_0$ is that the matrix:

\begin{equation}\label{NLkal}
O(\textbf{x}_0,\textbf{u}^*)=
\begin{bmatrix}
& \frac{\partial{L_\textbf{f}^0 h_1}}{\partial x_1} & \frac{\partial{L_\textbf{f}^0 h_1}}{\partial x_2} & ... & \frac{\partial{L_\textbf{f}^0 h_1}}{\partial x_N} \\
& ... & ... & ... & ... \\
& \frac{\partial{L_\textbf{f}^0 h_M}}{\partial x_1} & \frac{\partial{L_\textbf{f}^0 h_M}}{\partial x_2} & ... & \frac{\partial{L_\textbf{f}^0 h_M}}{\partial x_N} \\
& \vdots & \vdots & \vdots & \vdots \\
& \frac{\partial{L_\textbf{f}^{N-1} h_1}}{\partial x_1} & \frac{\partial{L_\textbf{f}^{N-1} h_1}}{\partial x_2} & ... & \frac{\partial{L_\textbf{f}^{N-1} h_1}}{\partial x_N} \\
& ... & ... & ... & ... \\
& \frac{\partial{L_\textbf{f}^{N-1} h_M}}{\partial x_1} & \frac{\partial{L_\textbf{f}^{N-1} h_M}}{\partial x_2} & ... & \frac{\partial{L_\textbf{f}^{N-1} h_M}}{\partial x_N} \\
\end{bmatrix}
\end{equation}
is of rank $N$. If the system is locally observable about any $\textbf{x}_0\in R^N$, the system is said to be  observable. Note that testing observability may require that the rank observability test is repeated for potentially infinitely many points $\textbf{x}_0$. Alternatively, one may recur to using a symbolic method for calculating the rank of (7).
\end{definition}

\begin{definition}{\textit{Observability of Linear System.}}
When a system is linear time invariant \cite{kalman1963mathematical},
\begin{equation}\label{Lsys}
	\begin{aligned}
	\dot{\textbf{x}}(t) & = A\textbf{x}+B\textbf{u}\\
	\textbf{y}(t) & = C\textbf{x}(t)
	\end{aligned}
\end{equation}
the condition for observability in Eq.\ \eqref{NLkal} reduces to the Kalman rank condition, i.e., the matrix
\begin{equation*}
K = \left[C^T,(CA)^T,\cdots,(CA^{N-1})^T\right]^T 
\end{equation*}
must have rank $N$. In the case that Eq.\ \eqref{Lsys} represents the dynamics of a dynamical network,  the matrix $A = \{A_{ij}\}$ is called the adjacency matrix. 
\end{definition}

\begin{definition}{\textit{Network Symmetries.}}
	We consider the group of network symmetries (or automorphisms of a graph). Each symmetry is represented by a permutation matrix that permutes with the adjacency matrix $A$, i.e., $PA=AP$. Given a matrix $A$ the set of all the network symmetries can be computed by using available efficient discrete algebra routines \cite{sage}.
\end{definition}

\section{\textbf{A Graphical Approach to Observability}}
The algebraic methods introduced in Sec.\ II to determine observability of a nonlinear system can get quite complicated as $N$, the number of states, increases. In \cite{liu2013observability} a graphical approach is proposed, which  reduces to the analysis of an inference diagram (ID) that describes the interactions between the state variables. This approach presents computational advantages. The \emph{inference diagram} is a graphical representation of the interactions between state variables: consider two variables $x_i$ and $x_j$, if variable $x_i$ appears in the dynamical equation for the evolution of variable $x_j$ ($x_i$ affects $x_j$), then there will be a directed edge from state $x_i$ to state $x_j$ in the inference diagram. Note here that the convention in this paper to construct the   ID  is different from that in Ref. \cite{liu2013observability}, namely if state $x_i$ affects state $x_j$ we draw a directed edge from $x_i$ to $x_j$ in the   ID  (while in \cite{liu2013observability} we would have a directed edge from $x_j$ to $x_i$). 
\begin{figure}[!ht]
	\centering
	\includegraphics[scale=0.6]{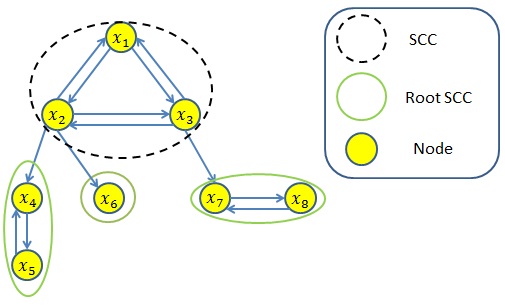}
	\caption{Inference diagram and its partition in strongly connected components. The   ID  is used to identify strongly connected components (SCCs) and root strongly connected components (root SCCs). In the figure (non-root) SCCs are delimited by dashed ovals and root SCCs by solid ovals. In order to achieve observability, at least one node needs to be selected from each root SCC.}
	\label{fig:example}
\end{figure} 

The   ID  contains information that can be used to identify the sensor nodes required to achieve observability.
Consider a directed graph $G = (V ; E)$, where $V$ is the set of nodes and $E\subseteq V \times V$ is the set of directed edges. A path from node $v_0$ to node $v_k$ in $G$ is an alternating sequence of the type $(v_0, (v_0, v_1), v_1, (v_1, v_2), ..., v_{k1}, (v_{k1},v_k), v_k)$ of nodes and edges that belong to $V$ and $E$, respectively. The entries of the adjacency matrix  $A$ are  $A_{ij} = 1 $, if $(v_j,v_i)\in E$ and $A_{ij} = 0$, if  $(v_j,v_i)\notin E$.  Two nodes $v$ and $w$ in $G$ are path equivalent if there is a path from $v$ to $w$ and a path from $w$ to $v$. Path equivalence partitions $V$ into maximal disjoint sets of path equivalent nodes. These sets are called the \textit{strongly connected components} (SCCs) of the graph \cite{Nuutila94onfinding}.
This means that, in a SCC, there exists a path from any node to all other nodes in that SCC. The sensor nodes predicted from the GA analysis are chosen from the root SCCs. A root SCC is a SCC which has no outgoing links, meaning that their information cannot be concluded from any other node in the inference diagram. However, root SCCs do have incoming links from other SCCs. This implies that information of nodes within other SCCs, can be inferred from the root SCCs. In order to achieve observability it is necessary that at least one sensor node (any node) is selected from each root SCC \cite{liu2013observability}. In the case presented in Fig.\ref{fig:example}, since there are three root SCCs, three sensor nodes must  be selected to achieve observability. Note that if the diagram is undirected and connected, choosing any node in the graph as sensor node would provide observability.

When applied to the   ID  of a network of coupled dynamical systems, GA may be used to identify a set of  \textit{sensor nodes} \cite{liu2013observability} to be monitored to yield observability.
In what follows, we briefly review the main results from \cite{liu2013observability}:

R1) The set of sensor nodes identified by the GA are necessary  for observing any linear or nonlinear dynamical system.

R2) If the dynamics is linear the maximal matching method introduced in \cite{liu2011controllability} returns the set of nodes that are necessary, but also the minimum set sufficient for observability.

R3) For nonlinear systems, under general  assumptions, the sensor set predicted by GA is not only necessary but also sufficient (for more details the reader is directed to the supplementary information of Ref.\ \cite{liu2013observability}).

Result R3) is supported in \cite{liu2013observability} by using a mix of the structural controllability  analysis \cite{lin1974structural} and an extensive numerical analysis. The  sufficiency argument is based on the assumptions that ($i$) all the nodes are accessible from the sensors which is guaranteed by placing the sensors in the root SCCs and ($ii$)  for each node $i$ $\partial f_i/\partial x_i \neq 0$, i.e., the dynamics of each node $i$ is affected by the state of the node itself.

\textit{Assumption.} Throughout  this paper we assume $\partial f_i/\partial x_i \neq 0$, $i=1,..,N$. Note that for the sake of simplicity, in the figures of this paper we omit to represent self-loops at each node. However, though not graphically shown, those self-connections are present.

We depart from result R3, which applies to general nonlinear networks, and extend it to consider the case where network symmetries are present. Our results are mainly meant to apply to large nonlinear complex networks in the presence of network symmetries.
Whether these symmetries are a generic feature or not of complex networks, it is a matter of discussion. While an analysis of model networks seems to indicate that network symmetries may sometimes emerge in small numbers in model networks \cite{NC}, it has also been reported that they are pervasive in the case of many real networks \cite{macarthur2007automorphism}. For example,  in Ref.\ \cite{macarthur2007automorphism} it was found that the Human B cell genetic interactions network is characterized
by approximately $5.9\times10^{13}$ symmetries, the C. elegans genetic interaction network is characterized by approximately $6.9\times10^{161}$ symmetries, and the network of the Internet at the Autonomous System
level by approximately $1.2\times10^{11298}$ symmetries (!) Hence, the presence of network symmetries appears to be a typical  feature of several real networks.

\section{\textbf{Symmetries and Observability}}

In this section we will analyze cases in which symmetries occur and their effect on observability. In what follows, we will present several inference diagrams with symmetries. For each individual diagram, we will use GA to determine SCCs as well as root SCCs and we will methodically examine the conditions that must be met for observability.

While we do not attempt to present an exhaustive treatment of the effects of symmetries on the observability of networks, we will present several examples and we will propose and show that: (\textit{i}) the GA method presented in the previous section fails in the presence of symmetries and (\textit{ii}) necessary conditions for observability can be restored by selecting appropriate additional sensor nodes. Our main goal in this section is to show that a GA can be conveniently exploited not only to find (root) SCCs, but also to detect symmetries. Given a network, available software is capable to efficiently provide all the network symmetries, e.g. SAGE \cite{sage}. Here we propose that SCCs analysis in conjunction with the analysis of symmetries in the   ID  may allow us to identify the necessary set of sensor nodes to achieve observability.

\begin{figure}[!ht]
	\centering
	\includegraphics[scale=0.6]{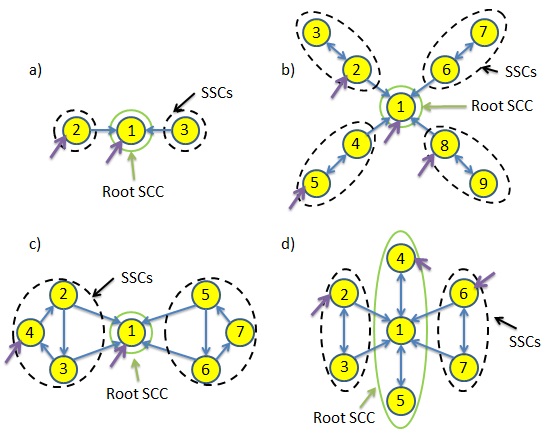}
	\caption{Inference diagram representation of four networks displaying symmetries. SCCs and root SCCs are delimited by dashed ovals and solid ovals, respectively. For each network, a minimal set of sensor nodes to achieve observability is determined (the sensor nodes are those pointed by small purple arrows).}
	\label{fig:sym1}
\end{figure} 

In what follows, we will specify the network nonlinear equations (1) and (2) to be,
\begin{equation}\label{eq9}
	\begin{aligned}
	\dot{x}_i(t) & =  g(x_i,A_{i1} q(x_1), A_{i2} q(x_2),..., A_{iN} q(x_N)) \\
	\textbf{y}(t) & =C \textbf{x}(t) 
	\end{aligned}
\end{equation}
where $q(x_i(t))$ is the scaler output function of node $i$, $i=1,\cdots,N$, and $ g: \mathbb{R}\times \underbrace{\mathbb{R} \times... \times \mathbb{R}}_N \rightarrow \mathbb{R}$ is a symmetric function of its variables.

\begin{definition}
	 A  symmetry for  the set of equations (9)  is a permutation between tuples of nodes for which the dynamics  is equivariant.
\end{definition}

It can be easily shown that a network symmetry  (Definition 3) yields an equation symmetry (Definition 4) for Eq.\ (9).
As an example, we consider the following nonlinear dynamic equations for the network in Fig.\ \ref{fig:sym1}a,
\begin{equation*}
\begin{aligned}
\dot{x}_1 & = -kx_1 x_2 x_3\\
\dot{x}_2 & = -kx_2 x_1 \\
\dot{x}_3 & = -kx_3 x_1\\
\end{aligned}	
\end{equation*}
The GA method predicts node 1 is the only sensor node. By monitoring $y=x_1$, we check the observability of the nonlinear system using the algebraic method described in Sec.\ II. The Lie derivatives of $y$ are:
\begin{equation*}
\begin{aligned}
\Phi_1^{(0)} & = L_\textbf{f}^0y = x_1\\
\Phi_1^{(1)} & = L_\textbf{f}^1y = \dot{x}_1  = -kx_1 x_2 x_3\\
\Phi_1^{(2)} & = L_\textbf{f}^2y = k^2(2 x_1^2x_2x_3 + x_1x_2^2x_3^2)\\
\end{aligned}	
\end{equation*}
For arbitrary $\textbf{x}$ and the constant $k$, the observability matrix,
\begin{equation*}
O = 
\begin{bmatrix}
& 1 & 0 &  0\\
& O_{21}& O_{22} &  O_{23} \\
& O_{31} &  O_{32} & O_{33}\\
\end{bmatrix}
\end{equation*}
where, $O_{21} = -kx_2x_3$, $O_{22}=-kx_1x_3$, $O_{23}=-kx_1x_2$, $O_{31} = k^2(x_2^2x_3^2 + 4x_1x_2x_3)$, $ O_{32}= 2k^2(x_1^2x_3 + x_2x_1x_3^2) $, $O_{33} =  2k^2(x_1^2x_2 + x_3 x_1 x_2^2) $. A symbolic rank calculation shows that $\mbox{rank}(O) = 2 < 3$, i.e, the system is not observable by only sensing node 1. Note that because the entries of the matrix $O$ are not independent of each other, the theory of structural controllability and observability \cite{lin1974structural} does \emph{not} exactly apply to this case \cite{liu2013observability}.

Now we consider the case that  we choose node 1 and 2 as sensor nodes. By monitoring $\textbf{y} = [x_1  \quad  x_2]^T$, we check the observability of the system. The Lie derivatives of $\textbf{y}$ are:
\begin{equation*}
\begin{aligned}
\Phi_1^{(0)} & = L_\textbf{f}^0h_1 = x_1\\
\Phi_1^{(1)} & = L_\textbf{f}^1h_1 = \dot{x}_1  = -kx_1 x_2 x_3\\
\Phi_1^{(2)} & = L_\textbf{f}^2h_1 = k^2(2 x_1^2x_2x_3 + x_1x_2^2x_3^2)\\
\Phi_2^{(0)} & = L_\textbf{f}^0h_2 = x_2\\
\Phi_2^{(1)} & = L_\textbf{f}^1h_2 = \dot{x}_2  = -kx_1 x_2\\
\Phi_2^{(2)} & = L_\textbf{f}^2h_2 = k^2(x_1^2 x_2 + x_3 x_1 x_2^2)\\
\end{aligned}	
\end{equation*}
For arbitrary $\textbf{x}$ and the constant $k$, the observability matrix,
\begin{equation*}
O = 
\begin{bmatrix}
& 1 & 0 &  0\\
& 0 & 1 &  0\\
& O_{31} &  O_{32} & O_{33}\\
& O_{41} &  O_{42} & O_{43}\\
& O_{51} &  O_{52} & O_{53}\\
& O_{61} &  O_{62} & O_{63}\\
\end{bmatrix}
\end{equation*}
where, $O_{31} = -kx_2x_3$, $O_{32}=-kx_1x_3$, $O_{33}=-kx_1x_2$, $O_{41} = -k x_2$, $ O_{42}= - k x_1 $, $O_{43} =  -k x_1 x_2$, $O_{51} = k^2(x_2^2x_3^2 + 4x_1x_2x_3)$, $ O_{52}= 2k^2(x_1^2x_3 + x_2x_1x_3^2) $, $O_{53} =  2k^2(x_1^2x_2 + x_3 x_1 x_2^2) $, $O_{61} = k^2(x_3 x_2^2 + 2 x_1  x_2)$, $ O_{62}= k^2(x_1^2 + 2 x_2 x_3 x_1) $, $O_{63} =  k^2 x_1 x_2^2$. It can be shown  using symbolic rank calculation that $\mbox{rank}(O) = 3$, i.e, the system is observable by sensing nodes 1 and 2.

From Fig.\ \ref{fig:sym1}a we see that swapping nodes 2 and 3 does not affect the dynamics of node 1. In Fig.\ \ref{fig:sym1}b, any permutation of the ordered pair of nodes (2,3), (4,5), (6,7) and (8,9) leaves the dynamics of node 1 unaltered. In Fig.\ \ref{fig:sym1}c, a permutation of the ordered triplets of the nodes (2,3,4) and (5,6,7) leaves the dynamics of node 1 unaltered. Fig.\ \ref{fig:sym1}d, shows four independent symmetries occurring in the same diagram, namely, swapping nodes 4 and 5 would leave the dynamics of the rest of the network unaltered, as well as swapping nodes 2 and 3, and 6 and 7. Moreover, another symmetry is generated by the pairs (2,3) and (6,7).
Hereafter, we will discuss how the symmetries of the diagrams in Fig.\ \ref{fig:sym1}  limit their observability and how additional sensor nodes can be selected to enable it. In Fig.\  \ref{fig:sym1}a, one can see a network composed of three nodes 1,2, and 3, where node 1 forms a root SCC and nodes 2 and 3 form individual SCCs. Moreover, nodes 2 and 3 are symmetric. This means that monitoring node 1 alone is not sufficient to achieve observability, while monitoring node 1 and either node 2 or 3 will provide observability.
In Fig.\ \ref{fig:sym1}b, the symmetric ordered pairs (2,3), (4,5), (6,7), (8,9) form individual SCCs. To achieve observability we need to monitor node 1 along with at least three other nodes and each of these extra nodes need to be selected from three different symmetric pairs. The third case shown in Fig.\ref{fig:sym1}c is that of a network in which the ordered triples (2,3,4) and (5,6,7) are symmetric. Similarly to the case presented in Fig.\ \ref{fig:sym1}b, for observability, we need to monitor node 1 and at least one other node from either ordered triplet. In the forth case, Fig.\ \ref{fig:sym1}d, there are symmetries both inside the root SCC, as well as in the other two SCCs. A necessary set of sensor nodes for this case will be: either node 4 or 5, either node 2 or 3 and either node 6 or 7. An algorithm to identify the minimal set of necessary sensor nodes for observability in the presence of symmetries will be presented in the Sec.\ V.
\begin{figure}[!ht]
		\centering
	\includegraphics[scale=0.65]{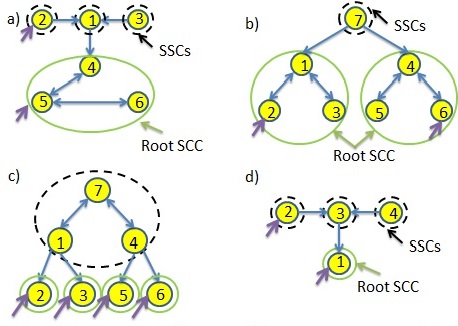}
	\caption{a) Inference diagram representation of a network with symmetries between two SCCs. b) Inference diagram representation of a network with symmetries between two root SCCs. c) Inference diagram representation of a network with symmetries between the SSCs. d) Inference diagram representation of a network with symmetries between two SCCs.}
	\label{fig:sym2}
\end{figure}

We continue to analyze cases in which symmetries occur. We see from Fig.\ \ref{fig:sym2}a, that nodes 2 and 3, are symmetric with respect to any node in the root SCC (if they are symmetric with respect to one they are symmetric with respect to all). In this case, for observability, we need to monitor either one of the nodes in the root SCC along with either node 2 or node 3. The case in Fig.\ \ref{fig:sym2}a is that of a symmetry between two nodes in two different SCCs. By definition of SCC and root SCC, a node in a  SCC and a node in a root SCC cannot be symmetric. However, there can be symmetries between nodes that belong to the same root SCC and also between nodes in different root SCCs (as shown in Fig.\ \ref{fig:sym2}b). In Fig.\ \ref{fig:sym2}b, the ordered triplets (1,2,3) and (4,5,6) are symmetric and there exist symmetries within each root SCC, nodes 2 and 3 are symmetric as well as nodes 5 and 6. In this case, by monitoring a node from each root SCC, the symmetries inside each root SCC are resolved and observability is achieved. The same symmetries are present in Fig.\ \ref{fig:sym2}c, but in this case the triplets (1,2,3) and (4,5,6) contain nodes from both SCCs and root SCCs. However, nodes belonging to non-root (root) SCCs can only be swapped with other nodes belonging to non-root (root) SCCs.

In Fig.\ \ref{fig:sym2}d, nodes 2 and 4 are symmetric. Hence, as in the previous cases, monitoring only the root SCC is not sufficient for observability. Because of the symmetry generated by nodes 2 and 4, we need to monitor the root SCC node 1 and either node 2 or 4.

\section{Algorithm to Determine the Minimal Set of Necessary Sensor Nodes for Observability}
In the cases presented in Figs.\ \ref{fig:sym1} and \ref{fig:sym2},  in which symmetries are present, monitoring one node from each root SCC is not sufficient to obtain observability. In general, additional nodes need to be monitored. Here we focus on the effect of the network symmetries and  present an algorithm that, given a   ID  with symmetries, returns a minimal set of necessary sensor nodes  $\mathcal{S}$ for  observability. First we present the algorithm for the case that there are no symmetries. Consider a diagram with $z$ root SCCs. Each root SCC can be described by a set as follows,
\begin{equation}
	\begin{aligned}
		\Pi^1& =[\Pi^1_1,\Pi^1_2,\cdots,\Pi^1_{n_1}],\\
		\Pi^2& =[\Pi^2_1,\Pi^2_2,\cdots,\Pi^2_{n_2}], \quad \cdots\\
		\Pi^z& =[\Pi^z_1,\Pi^z_2,\cdots,\Pi^z_{n_z}]
	\end{aligned}
\end{equation}
where $n_i$ is the number of elements in the set $i=1,...,z$. Then as we need to choose a node from each root SCC, and as SCCs are disjoint sets, the minimal set of necessary sensor nodes for observability $\mathcal{S}=[\Pi^1_*,\Pi^2_*,...,\Pi^z_*]$, where $\Pi^i_*$ is any element from the set $\Pi^i$ \cite{liu2013observability}.

In the case of a   ID  with $c>0$ symmetries, let a symmetry $S_i$ be represented by the following list of $m_i$ $l_i$-tuples,
\begin{equation}
S_i=((t_i^1),(t_i^2),...,(t_i^{m_i})),
\end{equation}
$i=1,...,c$. Here, $m_i$ is the total number of tuples in $S_i$ (note that any permutation of  tuples belonging to the same symmetry  is equivariant with respect to the network dynamics (9)) and $l_i$ is the number of elements of the tuples in $S_i$. We call $m_i$ the \emph{number} of symmetry $S_i$ and $l_i$ the \emph{length} of symmetry $S_i$. Without loss of generality, we label the symmetries so that $l_1 \leq l_2 \leq ... \leq l_c$, \emph{i.e.}, in order of nondecreasing length.

Now we introduce the definition of \emph{internal} and \emph{external} symmetry. A symmetry $S_i$ is internal to another symmetry $S_j$, $j \neq i$ (or, equivalently, $S_j$ is external to $S_i$) if there exists a tuple of $S_j$, say $t_j^k$, such that all the entries of the tuples $t_i^\ell$ are also entries of $t_j^k$, $\ell=1,...,m_i$.

In order to achieve observability, we need to include in $\mathcal{S}$: (i) at least one element from each set $\Pi_i$, $i=1,...,z$ and (ii) at least one element from $(m_i-1)$ different tuples of $S_i$, $t_i^1,t_i^2,...,t_i^{m_i}$, $i=1,...,c$. Given a set of sensor nodes $\mathcal{S}$, if condition (ii) is verified for a given $i$, we say that the root SCC $\Pi_i$ is resolved, and if condition (ii) is verified for a given $i$, we say that symmetry $S_i$ is resolved.

We now present an efficient algorithm to obtain a minimal set of necessary sensor nodes $\mathcal{S}$ for  observability, in the presence of symmetries. Our algorithm is based on the observation  that if a symmetry does not present internal symmetries, it will be resolved by choosing any $(m_i-1)$ elements from $(m_i-1)$ different tuples. Otherwise, if a symmetry presents internal symmetries, then resolving the internal symmetries ensures that the external symmetry is resolved as well. As the algorithm proceeds by considering symmetries of nondecreasing length, it first resolves internal symmetries, which automatically also resolves the external ones.

The algorithm consists of the following steps:

\begin{enumerate}
	\item Consider symmetries in the order of nondecreasing length. For each symmetry that has not been already resolved, select any $(m_i-1)$ elements from $(m_i-1)$ different tuples and add them to $\mathcal{S}$. Terminate when all symmetries have been resolved.
	\item Consider one by one all the root strongly connected components. For each root strongly connected component that has not been already resolved, select any element and add it to $\mathcal{S}$.
\end{enumerate}

\begin{figure}[!ht]
	\centering
	\includegraphics[width=0.45\textwidth]{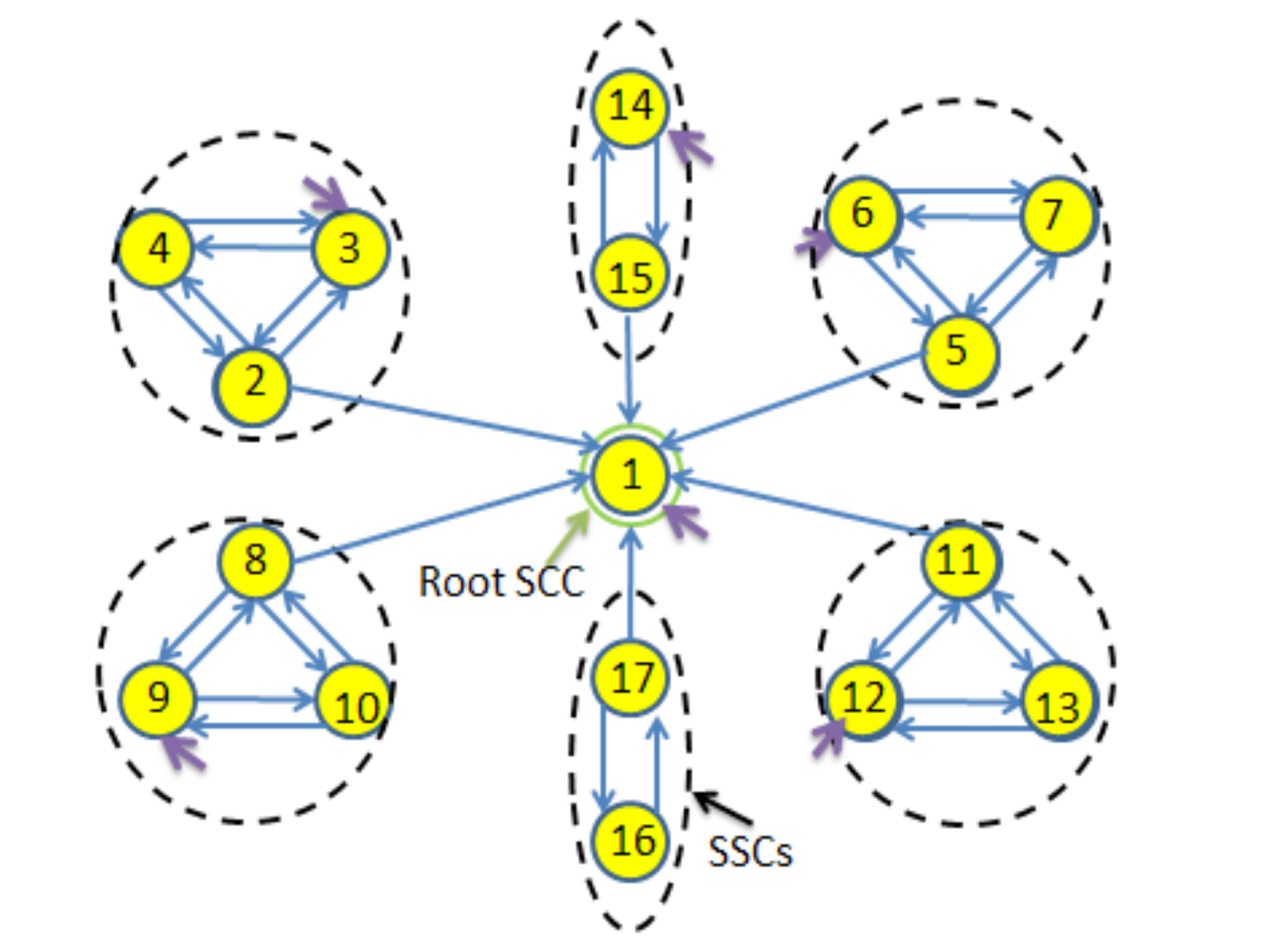}
	\caption{The algorithm described in the text to obtain $\mathcal{S}$, the minimal set of nodes for observability, has been applied to the network in the figure.}
	\label{fig:sym3}
\end{figure}

As an example, we apply our algorithm to the   ID  representation in Fig.\ \ref{fig:sym3}. This network is directed and is composed of 6 non-root SCCs  and 1 root SCC.  We computed the network symmetries by using the \texttt{SAGE} routine \texttt{AUTOMORPHISM\_GROUP}.
The symmetries for this network are given by $S_1=((3),(4))$, $S_2=((6),(7))$, $S_3=((9),(10))$, $S_4=((12),(13))$, $S_5=((14,15),(16,17))$, and $S_6=((2,3,4), (5,6,7), (8,9,10), (11,12,13))$. In this case, $m_1=m_2=m_3=m_4=m_5=2$ and $m_6=4$; moreover, $l_1=l_2=l_3=l_4=1$, $l_5=2$, and $l_6=3$. Note that the symmetries $S_1,S_2,S_3$, and $S_4$ are all internal to $S_6$. According to our algorithm, we resolve the internal symmetries first. To do that, we choose one node from each one of the symmetries $S_1,S_2,S_3$, and $S_4$ and add them to $\mathcal{S}$. For example, a possible choice is to set $\mathcal{S}=[3,6,9,12]$. We see that with this $\mathcal{S}$, we have automatically resolved the external symmetry $S_6$. In fact, $\mathcal{S}$ contains $(m_6-1)=3$ different nodes from $(m_6-1)=3$ different tuples of $S_6$. We then need to resolve $S_5$ by choosing one node from either the tuple $(14,15)$ or  $(16,17)$. Say we choose $14$, then the updated set $\mathcal{S}$ is $[3,6,9,12,14]$.  The network has one root SCC. According to step 2 of the algorithm, we need to choose one node from this root SCC, unless one is already included in $\mathcal{S}$. Here node 1 is the only option and it has not been already selected. Then, the final set is $\mathcal{S}=[1,3,6,9,12,14]$. Note that this set is minimal, as eliminating any node from $\mathcal{S}$ would correspond to loss of observability.

\begin{figure}[!ht]
		\centering
		\includegraphics[width=0.48\textwidth]{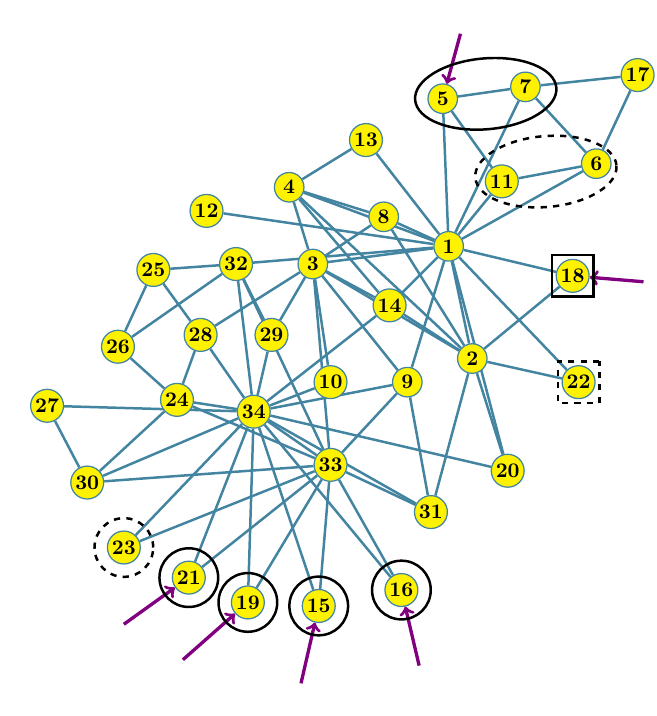}
		\caption{Undirected network for Zachary's Karate Club \cite{zachary1977information,girvan2002community}. The ovals, circles and rectangles are presenting symmetries.}
		\label{fig:real1}
\end{figure}

We now apply our algorithm to two different real networks. First we consider the Zachary's Karate club which represents friendship relations among 34 members \cite{zachary1977information,girvan2002community}. Since the network is undirected and connected, the whole network coincides with a root SCC (see Fig.\ \ref{fig:real1} ). The symmetries for this network have been evidenced in Fig.\ \ref{fig:real1}, where same shapes (ovals, rectangles and circles) have been used to indicate nodes that  are symmetric. By applying the algorithm, we find a minimum set of nodes necessary for observability $\mathcal{S}=[5 ,   15  ,  16  ,  18    ,19  , 21]$.

We now consider the Caenorhabditis Elegans neuronal networks \cite{varshney2011structural} formed of $279$ neurons. These can be linked through connections of two different types, either chemical synapses or gap junctions.   Here we focus on the network formed of chemical synapses alone, which is directed. This network is not fully connected, it has 42 SCCs, 11 of which are root SCCs. This network also presents some symmetries. A minimal set of necessary sensor nodes $\mathcal{S}$ for  observability of the chemical synapses network can be found by using our algorithm and the results of our computations are presented in Table I.
\begin{table}[!h]
	\centering
	\begin{tabular}{|c|c|c|}
		\hline
		No. of nodes & Name of The Neurons & Neuron Label No. \\ \hline
		1            & IL2DL               & 1                     \\ \hline
		2            & IL2DR               & 6                     \\ \hline
		3            & ASIL                & 76                    \\ \hline
		4            & ASIR                & 90                    \\ \hline
		5            & AINL                & 120                   \\ \hline
		6            & SDQR                & 179                   \\ \hline
		7            & DB05                & 204                   \\ \hline
		8            & AS08                & 220                   \\ \hline
		9            & PVDR                & 227                   \\ \hline
		10           & DVB                 & 253                   \\ \hline
		11           & PLNR                & 260                   \\ \hline
		12           & PHCR                & 274                   \\ \hline
		13           & PLML                & 279                   \\ \hline
	\end{tabular}
	\caption{The minimal set of necessary sensor nodes for observability of chemical synapse network of C. elegans(directed network).}
	\label{ex:table1}
\end{table}

\section{\textbf{Conclusions}}

In this paper we have presented an efficient algorithm to determine a minimal set of necessary sensor nodes for observability of complex network in the presence of network symmetries. These symmetries have been shown to be a typical property of several real networks \cite{macarthur2007automorphism}.  We have shown that  observability requires  by a proper selection of the sensor nodes, which takes into account the network symmetries.

Our method based on a graphical approach could be used to uncover the conditions required for the observation of neural and genetic networks, for which access to all the network states is limited and knowledge about the connectivity pattern is crucial. For example, this is relevant to the estimation of neuronal \cite{white1986structure,sporns2005human} and genetic networks \cite{de2002linking,friedman2004inferring,de2006unravelling,faith2007large,marbach2009replaying,de2010advantages,marbach2010revealing} and to numerous engineering applications \cite{autariello2013estimating,bezzodecentralized,husseinbayesian,sorrentino2008adaptive,sorrentino2009using}.


\end{document}